\documentclass{article}%
\usepackage{amsmath}
\usepackage{amsfonts}
\usepackage{amssymb}
\usepackage{graphicx}%
\begin{document}

\title{The Effects of Non Standard Neutrino Interactions (NSIs) in $K_{L}%
^{0}\rightarrow\pi^{0}\nu\overline{\nu},B^{+}\rightarrow\pi^{+}\nu
\overline{\nu},B^{+}\rightarrow K^{\ast+}\nu\overline{\nu}$ and $B\rightarrow
X_{s}\nu\overline{\nu}$}
\author{Shakeel Mahmood$^{{\footnotesize (1)}}$; Farida Tahir; Azeem Mir\\\textit{Comsats Institute of Information Technology,}\\\textit{\ Department of Physics, Park Road,\ Chek Shazad,Islamabad}\\$^{({\footnotesize 1)}}${\footnotesize shakeel\_mahmood@hotmail.com}}
\date{}
\maketitle

\begin{abstract}
We study the rare decays $K_{L}^{0}\rightarrow\pi^{0}\nu\overline{\nu},$
$B^{+}\rightarrow\pi^{+}\nu\overline{\nu},B^{+}\rightarrow K^{\ast+}%
\nu\overline{\nu}$~and $B\rightarrow X_{s}\nu\overline{\nu}$ for the search of
NSIs. We want to constraint the NSIs by using these reactions. We show that
there is a strong dependence of these reactions on new physics free parameter
$\epsilon_{\tau\tau}^{QL},$where $Q=u,c,t$. We include second and third
generation of quarks in the loop for these decays. We show that the $K_{L}%
^{0}\rightarrow\pi^{0}\nu\overline{\nu}$ is providing very precise bounds as
compared to all other semileptonic decays, having neutrinos in their final
state. We further show that the interference between standard model and NSIs
is giving dominant contribution for $B^{+}\rightarrow\pi^{+}\nu\overline{\nu
},B^{+}\rightarrow K^{\ast+}\nu\overline{\nu}$~and $B\rightarrow X_{s}%
\nu\overline{\nu}$. We point out that the constraints for $\epsilon_{\tau\tau
}^{cL}$ and $\epsilon_{\tau\tau}^{tL}$ are more precise as compared to
$\epsilon_{\tau\tau}^{uL}$. The analysis of $B^{+}\rightarrow\pi^{+}%
\nu\overline{\nu},B^{+}\rightarrow K^{\ast+}\nu\overline{\nu}$ and
$B\rightarrow X_{s}\nu\overline{\nu}$, provide us that the $u$ quark induced
Br of NSIs are giving very very small contribution. We also compare these
decays to the decays of charm and kaons having neutrinos in the final state.

%


%


\end{abstract}

\section{Introduction}

After the remarkable discovery of Higgs by the ATLAS \cite{Atlas} and CMS
\cite{CMS} collaboration and confirmation in \cite{SM higgs} that it is
Standard Model (SM) higgs, one important question arises. Is there any room
for new physics (NP) beyond SM? No doubt SM predictions have been verified
experimentally to the highest level of precision \cite{Supersymmetry}%
\cite{Supersymmetry 1}. But, along with other limitations, SM lacks any
explanation for a possible pattern for particle mass, known as mass hierarchy
problem. SM can not predict top quark mass without experimental evidence. The
experiments on B meson \cite{M. Tanaka Z. Phys. C}\cite{J. P. Lees et
al.}\cite{J. P. Lees} are also giving some cracks in standard model. We are
yet unable to explain dark matter and matter anti-matter asymmetry. Gravity is
not included in the SM. Theoretically SM is thought to be unsatisfactory and
there can be some new particles as well as new interactions. It has been
believed that standard model is a low energy approximation of more general
theory. So, many theoretical extensions of SM has been presented. But, so far,
the only concrete evidence against it has been provided by the neutrino
oscillations \cite{Super-Kamiokande} \cite{Super-Kamiokande 1}%
\cite{Super-Kamiokande 2}\cite{Super-Kamiokande 3}\cite{Super-Kamiokande
4}\cite{Super-Kamiokande 5}\cite{Super-Kamiokande 6}. To explore NP the study
of mesonic rare decays involving neutrinos in final state, can be interesting.
These decays proceeds through flavor changing neutral currents (FCNC), highly
suppressed \cite{Alejandro Jaramillo} due to GIM mechanism \cite{S. L. Glashow
J} and occur at loop level \cite{Azeem}\cite{Takeo Inami and C.S. Lim}. The
discrepancies between experiments and theory (SM) for such reactions provide
us an excellent window towards NP. New particles can be added in the loops to
improve theory or we can have new interactions. So, FCNC reactions involving
neutrinos in the final state can be interesting.

Theoretically, $B^{+}\rightarrow\pi^{+}\nu\overline{\nu}$ and $K_{L}%
^{0}\rightarrow\pi^{0}\nu\overline{\nu}$ thought to be more clean than
$K^{+}\rightarrow\pi^{+}\nu\overline{\nu}~$because it has only top quark
contribution and no contribution from charm sector. Non standard neutrino
interactions (NSIs) of $K^{+}\rightarrow\pi^{+}\nu\overline{\nu}$ and
$D_{s}^{+}\rightarrow D^{+}\nu\overline{\nu}$ are studied in \cite{Chuan-Hung
Chen 2007} and \cite{shakeel} respectively and constrained are found for
$\epsilon_{\tau\tau}^{uL}$. The NSIs constraints for three generations of
quarks are given in \cite{shakeel 1}\ As all of these have same loop
structure, so, similar thing should happen to $B^{+}\rightarrow\pi^{+}%
\nu\overline{\nu}$ and $K_{L}^{0}\rightarrow\pi^{0}\nu\overline{\nu}$. Two
more loop level processes useful for the search of new physics are inclusive
$B\rightarrow X_{s}\nu\overline{\nu}$ and exclusive $B^{+}\rightarrow
K^{\ast+}\nu\overline{\nu}~$due to their theoretical cleanliness
\cite{Wolfgang Altmannshofer}.

In this paper we the scheme of study is as: first of all we\ give experimental
status of $B^{+}\rightarrow\pi^{+}\nu\overline{\nu},K_{L}^{0}\rightarrow
\pi^{0}\nu\overline{\nu},B^{+}\rightarrow K^{\ast+}\nu\overline{\nu}$ and
$B\rightarrow X_{s}\nu\overline{\nu}$ then we revise the SM contribution of
these reaction. Next, we study these reactions in NSIs with $u$ quark in the
loop which is the usual case of NSIs and obtain the constraints $\epsilon
_{\tau\tau}^{uL}$. Then we modify the operators for $c$ and $t$ quarks. and
find out the constraints $\epsilon_{\tau\tau}^{cL}$and $\epsilon_{\tau\tau
}^{tL}.$We compare these constraints among the three generations. We also
compare these constraints to the constraints of same type of reactions from D
and K decays. Then discussion and results are provided and conclusion is given
at the end.

\section{Experimental Status of\textbf{\ }$B^{+}\rightarrow\pi^{+}\nu
\overline{\nu}$\textbf{\ Decay }}

$B$ decays are being studied in the detectors like CLEO, CDF, BaBar, Belle,
ALEPH collaborations and LHCb but the decays involving neutrinos in the final
state will be tested at super $B$ factories in experimentally clean
environment. The detection of $B^{+}\rightarrow\pi^{+}\nu\overline{\nu}$ is a
hard task and currently we have only experimental bound for this reaction but
this will be in the range of super $B$ factories. The experimental bound is
given in the Table $1$ along$~$with$~$their SM prediction. Current
experimental bounds for this are $<9.8\times10^{-5}$but our experience with
$K^{+}\rightarrow\pi^{+}\nu\overline{\nu}$ guide us that the experimental
value should be of the order of $10^{-7}.$It means we are not discarding the
SM values but just searching for the small room for new physics.$~$%
Experimental value for $K_{L}^{0}\rightarrow\pi^{0}\nu\overline{\nu}$ is
$<2.6\times10^{-8}$\cite{PDG 2014}. $B^{+}\rightarrow K^{\ast+}\nu
\overline{\nu}$ is easy to measure experimentally because we have vector
particle in the final state, so polarization does matter and its latest bound
is $<4\times10^{-5}.$

Inclusive process $B\rightarrow X_{s}\nu\overline{\nu}$ are very difficult to
observe experimentally because we need to tag all the particles involve in
$X_{s}$ along with missing neutrinos. The limit available till to date is
$<64\times10^{-5}$, given in table 1 with reference.

\subsection{$M\rightarrow M^{/}\nu\overline{\nu}$ Decays in the Standard
Model}

Here mass of $M$
$>$
mass of $M^{/}$and both are representing the mesons. The SM calculation can be
divided into two categories, short distance and long distance..This can be
found from \cite{D. Rein and L. M. Sehgal} \cite{Lu} \cite{Rien} and
\cite{Andrzej J. Buras} that the dominant contribution for $M\rightarrow
M^{/}\nu\overline{\nu}$ comes from short distance because long distance
contribution is $10^{-3}$less than short distance, if the quark level process
is $b$ $\rightarrow d\nu\overline{\nu}$ or $s$ $\rightarrow d\nu\overline{\nu
}$. The quark level process for our decay is\ $b$ $\rightarrow d\nu
\overline{\nu}$ which can be represented by the feynman diagrams shown in
figure 1.%

\begin{figure}
\centering
%

{\includegraphics[
trim=0.000000in 0.000000in 0.062708in 0.065636in,
natheight=2.781200in,
natwidth=6.270800in,
height=2.7605in,
width=6.2733in
]%
{C: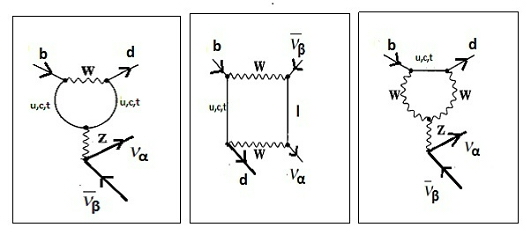}%
}
%

\caption{SM b decays to d netrino antinetrino}
\end{figure}%

In such reactions we can easily separate hadronic interactions from leptonic
interaction. For $B$ decays the dominant contribution comes from the short
distance just like $K$ decays and we use perturbation theory due to asymptotic
freedom. The effective Hamiltonian for such reactions quark level reaction
will be%

\[
H_{eff}^{SM}=\frac{G_{F}}{\sqrt{2}}\frac{\alpha_{em}}{2\pi\sin^{2}\theta_{W}%
}\underset{\alpha,\beta=e,\mu,\tau}{\Sigma}V_{tb}^{\ast}V_{td}X(x_{t}%
))\times(\overline{d}b)_{V-A}(\nu_{\alpha}\overline{\nu}_{\beta})_{V-A}%
\]

where $V-A$ in the subscript represents the vector and axial vector current
respectively. For such reactions charm quark contribution in the loop is
negligible in contrast to $K$ decay due to smallness of off diagonal $CKM$
element and $X(x_{t})$ is the loop integral of top-quark exchange \cite{Takeo
Inami and C.S. Lim}. For this reaction we have two penguin and one box diagram
\cite{G. Buchalla 1996} and sum of all give the contribution%

\[
X(x_{t})=\eta_{X}\frac{x_{t}}{8}[\frac{x_{t}+2}{x_{t}-1}+\frac{3x_{t}%
-6}{(x_{t}-1)^{2}}\ln x_{t}]
\]

Here $x_{t}=\frac{m_{t}^{2}}{m_{w}^{2}}$and $\eta_{X}=0.985$ is QCD small
distance correction. By using above Hamiltonian we can obtain Br as%

\[
Br(B^{+}\rightarrow\pi^{+}\nu\overline{\nu})_{SM}=r_{iso}\frac{3\alpha
_{em}^{2}}{|V_{ub}|^{2}2\pi^{2}\sin^{4}\theta_{W}}|V_{tb}^{\ast}V_{td}%
X(x_{t})|^{2}Br(B^{+}\longrightarrow\pi^{0}l^{+}\nu_{l})
\]
$r_{iso}\simeq0.94$ is the isospin breaking effect for B. It is discussed for
$K$ mesons in \cite{W. Marciano and Z. Parsa} which depends on at least three
things $(1)$ mass effect $(2)$ a suppression of about $4\%$ in neutral form
factor comes from $\eta-\pi$ mixing and $(3)$ \ about $2\%$ suppression due to
absence of log leading correction.

The reaction $B^{+}\longrightarrow K^{\ast+}\overline{\upsilon}\upsilon
,$proceed through quark level process $\overline{b}\rightarrow\overline{s}%
\nu\overline{\nu}$ and the effective Hamiltonian is same except to replace $d$
with $s$. Here we do not have a tree level reaction for normalization so we
have to use $B^{+}\longrightarrow\rho^{0}l^{+}\nu_{l}$%

\[
Br(B^{+}\rightarrow K^{\ast+}\nu\overline{\nu})_{SM}=r_{iso}\frac{3\alpha
_{em}^{2}}{|V_{ub}|^{2}2\pi^{2}\sin^{4}\theta_{W}}|V_{tb}^{\ast}V_{ts}%
X(x_{t})|^{2}Br(B^{+}\longrightarrow\rho^{0}l^{+}\nu_{l})
\]

For $B\rightarrow X_{s}\nu\overline{\nu}$ again the effective Hamiltonian is
same except to replace $d$ with $s$,and we do not have a tree level process
like $B^{+}\longrightarrow\pi^{0}l^{+}\nu_{l}$. So we have to normalize with
the process $B\rightarrow X_{c}\nu\overline{\nu}$ and due to different phase
spaces for $X_{s}$ and $X_{c},$we have to include other factors. The Br \ will be%

\[
Br(B\rightarrow X_{s}\nu\overline{\nu})_{SM}=\frac{3\alpha_{em}^{2}}{4\pi
^{2}\sin^{4}\theta_{W}}|\frac{V_{ts}V_{tb}X(x_{t})}{V_{cb}}|^{2}%
\frac{\overline{\eta}}{f(z)\kappa(z)}Br(B\longrightarrow X_{c}l\nu_{l})
\]

where $f(z)=1-8z+8z^{3}-z^{4}-12z^{2}\ln(z)$ with $z=\frac{m_{c}^{2}}%
{m_{b}^{2}}$

and $\kappa(z)=0.88$, $\eta=\kappa(0)=0.83$.

A useful discussion can about the factors can be found in \cite{Andrzej J.
Buras} and \cite{G. Buchalla 1996}. With the latest values of the constants we
have the Br%

\[
Br(B\rightarrow X_{c}\nu\overline{\nu})_{SM}=3.6\times10^{-5}%
\]
$K_{L}^{0}\rightarrow\pi^{0}\nu\overline{\nu}$ along with $K^{+}\rightarrow
\pi^{+}\nu\overline{\nu}$ is thought to be theoretically clean reactions for
the search of new physics. The effective Hamiltonian for $K_{L}^{0}%
\rightarrow\pi^{0}\nu\overline{\nu}$ can be written as%

\[
H_{eff}^{SM}=\frac{G_{F}}{\sqrt{2}}\frac{\alpha_{em}}{2\pi\sin^{2}\theta_{W}%
}\underset{\alpha,\beta=e,\mu,\tau}{\Sigma}V_{ts}^{\ast}V_{td}X(x_{t}%
))\times(\overline{d}s)_{V-A}(\nu_{\alpha}\overline{\nu}_{\beta})_{V-A}%
\]
and the Br can be extracted by normalizing with tree level reaction, so that
all the hadronic uncertainties are absorbed
\[
Br(K_{L}^{0}\rightarrow\pi^{0}\nu\overline{\nu})_{SM}=R_{iso}\frac
{3\alpha_{em}^{2}}{|V_{ub}|^{2}16\pi^{2}\sin^{4}\theta_{W}}\frac{\tau
(K_{L}^{0})}{\tau(K^{+})}|V_{ts}^{\ast}V_{td}X(x_{t})|^{2}Br(K^{+}%
\longrightarrow\pi^{0}e^{+}\nu_{e})
\]
Here isospin symmetry is exploited as
\[
\langle\pi^{0}|(\overline{d}s)_{V-A}|\overline{K}^{0}\rangle=\langle\pi
^{0}|(\overline{s}u)_{V-A}|K^{+}\rangle
\]
and isospin breaking effect is $R_{iso}=0.94$, and other values from \cite{PDG
2014} are used to obtained the Br
\[
Br(K_{L}^{0}\rightarrow\pi^{0}\nu\overline{\nu})_{SM}=2.06\times10^{-11}%
\]

\subsection{Model Independent Approach}

\ The NSI for the process is shown by the Fig 2%

\begin{figure}
\centering
%

{\includegraphics[
trim=0.000000in 0.000000in 0.000000in 0.189396in,
natheight=3.156600in,
natwidth=4.447700in,
height=3.0113in,
width=4.4996in
]%
{C: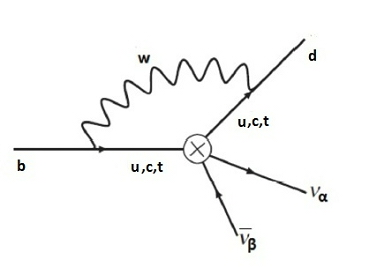}%
}

\bigskip%
\caption{NSIs b decays to d neutrino antineutrino}
\end{figure}%

and represented by%
\[
H_{eff}^{NSI}=\frac{G_{F}}{\sqrt{2}}(V_{tb}^{\ast}V_{tq}\frac{\alpha_{em}%
}{4\pi\sin^{2}\theta_{W}}\epsilon_{\alpha\beta}^{uL}\ln\frac{\Lambda}{m_{w}%
})\times(\nu_{\alpha}\overline{\nu}_{\beta})_{V-A}(\overline{q}b)_{V-A}%
\]

We use $V_{ub}=(4.15\pm0.49)\times10^{-3}$ and $Br(B^{+}\longrightarrow
\pi^{\circ}l^{+}\nu_{l})=(7.78\pm0.28)\times10^{-5}$ to find out Br%

\[
Br(B^{+}\longrightarrow\pi^{+}\overline{\upsilon}\upsilon)_{NSI}=r_{iso}%
\frac{\alpha_{em}^{2}}{|V_{ub}|^{2}8\pi^{2}\sin^{4}\theta_{W}}|V_{ub}^{\ast
}V_{ud}~\epsilon_{\alpha\beta}^{uL}\ln\frac{\Lambda}{m_{w}}|^{2}%
Br(B^{+}\longrightarrow\pi^{0}l^{+}\nu_{l})
\]

Although the current experimental results of $B$ decays are narrowing the gape
between theory and experiments but when we will get more precise experimental
data than we will need more accurate theoretical results. With the assumption
that experiments will give us the value of 10$^{-7},$we can constraint the
NSIs from this reaction. As $\alpha$ and $\beta$\ can be any lepton we take
them as $\tau,$because for other leptons we have already more precise
constraints \cite{S Davidson}.

Similarly NSIs Br of $B^{+}\longrightarrow K^{\ast+}\overline{\upsilon
}\upsilon$ can be found as
\[
Br(B^{+}\longrightarrow K^{\ast+}\overline{\upsilon}\upsilon)_{NSI}%
=\frac{\alpha_{em}^{2}}{|V_{ub}|^{2}8\pi^{2}\sin^{4}\theta_{W}}|V_{ub}^{\ast
}V_{us}~\epsilon_{\alpha\beta}^{uL}\ln\frac{\Lambda}{m_{w}}|^{2}%
Br(B^{+}\longrightarrow\rho^{0}l^{+}\nu_{l})
\]

For $B\rightarrow X_{s}\nu\overline{\nu}$ NSIs Br%

\[
Br(B\rightarrow X_{c}\nu\overline{\nu})_{NSIs}=\frac{\alpha_{em}^{2}}%
{16\pi^{2}\sin^{4}\theta_{W}}|\frac{V_{us}V_{ub}}{V_{cb}}\epsilon_{\alpha
\beta}^{uL}\ln\frac{\Lambda}{m_{w}}|^{2}\frac{\overline{\eta}}{f(z)\kappa
(z)}Br(B\longrightarrow X_{c}l\nu_{l})
\]

\[
Br(K_{L}^{0}\longrightarrow\pi^{0}\overline{\upsilon}\upsilon)_{NSI}%
=r_{iso}\frac{\alpha_{em}^{2}}{|V_{ub}|^{2}16\pi^{2}\sin^{4}\theta_{W}}%
|V_{ub}^{\ast}V_{ud}~\epsilon_{\alpha\beta}^{uL}\ln\frac{\Lambda}{m_{w}}%
|^{2}Br(K^{+}\longrightarrow\pi^{0}e^{+}\nu_{e})
\]

\subsection{Inteference between Standard Model and NSIs}

For all above decays the dominant contribution is coming from the interference
between the SM and NSIs. The Br of interference are obtained as%
\[
Br(B^{+}\longrightarrow\pi^{+}\overline{\upsilon}\upsilon
)_{Int\operatorname{erf}erence}=r_{iso}\frac{2\alpha_{em}^{2}}{|V_{ub}%
|^{2}8\pi^{2}\sin^{4}\theta_{W}}|V_{tb}^{\ast}V_{td}X(x_{t})V_{ub}^{\ast
}V_{ud}~\epsilon_{\alpha\beta}^{uL}\ln\frac{\Lambda}{m_{w}}|Br(B^{+}%
\longrightarrow\pi^{0}l^{+}\nu_{l})
\]

\[
Br(B^{+}\longrightarrow K^{\ast+}\overline{\upsilon}\upsilon
)_{Int\operatorname{erf}erence}=r_{iso}\frac{2\alpha_{em}^{2}}{|V_{ub}%
|^{2}8\pi^{2}\sin^{4}\theta_{W}}|V_{tb}^{\ast}V_{ts}X(x_{t})V_{ub}^{\ast
}V_{us}~\epsilon_{\alpha\beta}^{uL}\ln\frac{\Lambda}{m_{w}}|Br(B^{+}%
\longrightarrow\rho^{0}l^{+}\nu_{l})
\]

\[
Br(B\rightarrow X_{c}\nu\overline{\nu})_{Int\operatorname{erf}erence}%
=\frac{\alpha_{em}^{2}}{16\pi^{2}\sin^{4}\theta_{W}|V_{cb}|}|V_{ts}%
V_{tb}X(x_{t})V_{us}V_{ub}\epsilon_{\alpha\beta}^{uL}\ln\frac{\Lambda}{m_{w}%
}|\frac{\overline{\eta}}{f(z)\kappa(z)}Br(B\longrightarrow X_{c}l\nu_{l})
\]

\[
Br(K_{L}^{0}\longrightarrow\pi^{0}\overline{\upsilon}\upsilon
)_{Int\operatorname{erf}erence}=r_{iso}\frac{2\alpha_{em}^{2}}{|V_{ub}%
|^{2}16\pi^{2}\sin^{4}\theta_{W}}|V_{ts}^{\ast}V_{td}X(x_{t})V_{ub}^{\ast
}V_{ud}~\epsilon_{\alpha\beta}^{uL}\ln\frac{\Lambda}{m_{w}}|Br(K^{+}%
\longrightarrow\pi^{0}e^{+}\nu_{e})
\]

The contribution from interference for $B$ decays is is so large as compared
to the NSIs that NSIs effects can easily be ignored. So that the bounds on the
constraints are obtained from the interference only. But for the $K_{L}^{0}$
$\ $the $Br(K_{L}^{0}\longrightarrow\pi^{0}\overline{\upsilon}\upsilon
)_{Int\operatorname{erf}erence}$ is $10^{-3}$less than the $Br(K_{L}%
^{0}\longrightarrow\pi^{0}\overline{\upsilon}\upsilon)_{NSIs}$, so the
contribution is ignored. The numerical values are given in table 1 and 2.

\subsubsection{c-quark in the Loop}

For c-quark we have to modify the operators as and obtain the constraints%
\[
Br(B^{+}\longrightarrow\pi^{+}\overline{\upsilon}\upsilon
)_{Int\operatorname{erf}erence}=r_{iso}\frac{2\alpha_{em}^{2}}{|V_{ub}%
|^{2}8\pi^{2}\sin^{4}\theta_{W}}|V_{tb}^{\ast}V_{td}X(x_{t})V_{cb}^{\ast
}V_{cd}~\epsilon_{\alpha\beta}^{uL}\ln\frac{\Lambda}{m_{w}}|Br(B^{+}%
\longrightarrow\pi^{0}l^{+}\nu_{l})
\]

\[
\epsilon_{\tau\tau}^{cL}\leq1.5
\]

\[
Br(B^{+}\longrightarrow K^{\ast+}\overline{\upsilon}\upsilon
)_{Int\operatorname{erf}erence}=r_{iso}\frac{2\alpha_{em}^{2}}{|V_{ub}%
|^{2}8\pi^{2}\sin^{4}\theta_{W}}|V_{tb}^{\ast}V_{ts}X(x_{t})V_{cb}^{\ast
}V_{cs}~\epsilon_{\alpha\beta}^{uL}\ln\frac{\Lambda}{m_{w}}|Br(B^{+}%
\longrightarrow\rho^{0}l^{+}\nu_{l})
\]

\[
\epsilon_{\tau\tau}^{cL}\leq1.5
\]

For $B\rightarrow X_{s}\nu\overline{\nu}$ the Br and constraints are
\[
Br(B\rightarrow X_{c}\nu\overline{\nu})_{Int\operatorname{erf}erence}%
=\frac{\alpha_{em}^{2}}{16\pi^{2}\sin^{4}\theta_{W}|V_{cb}|}|V_{ts}%
V_{tb}X(x_{t})V_{cs}V_{cb}\epsilon_{\alpha\beta}^{uL}\ln\frac{\Lambda}{m_{w}%
}|\frac{\overline{\eta}}{f(z)\kappa(z)}Br(B\longrightarrow X_{c}l\nu_{l})
\]

\[
\epsilon_{\tau\tau}^{cL}\leq0.8
\]

\[
Br(K_{L}^{0}\longrightarrow\pi^{0}\overline{\upsilon}\upsilon
)_{Int\operatorname{erf}erence}=r_{iso}\frac{2\alpha_{em}^{2}}{|V_{ub}%
|^{2}16\pi^{2}\sin^{4}\theta_{W}}|V_{ts}^{\ast}V_{td}X(x_{t})V_{cb}^{\ast
}V_{cd}~\epsilon_{\alpha\beta}^{cL}\ln\frac{\Lambda}{m_{w}}|Br(K^{+}%
\longrightarrow\pi^{0}e^{+}\nu_{e})
\]

\subsubsection{t-quark in the Loop}

With t-quark we have following operator and constraint
\[
Br(B^{+}\longrightarrow\pi^{+}\overline{\upsilon}\upsilon
)_{Int\operatorname{erf}erence}=r_{iso}\frac{2\alpha_{em}^{2}}{|V_{ub}%
|^{2}8\pi^{2}\sin^{4}\theta_{W}}|V_{tb}^{\ast}V_{td}X(x_{t})V_{tb}^{\ast
}V_{td}~\epsilon_{\alpha\beta}^{tL}\ln\frac{\Lambda}{m_{w}}|Br(B^{+}%
\longrightarrow\pi^{0}l^{+}\nu_{l})
\]

\[
\epsilon_{\tau\tau}^{tL}\leq1.5
\]

\[
Br(B^{+}\longrightarrow K^{\ast+}\overline{\upsilon}\upsilon
)_{Int\operatorname{erf}erence}=r_{iso}\frac{2\alpha_{em}^{2}}{|V_{ub}%
|^{2}8\pi^{2}\sin^{4}\theta_{W}}|V_{tb}^{\ast}V_{ts}X(x_{t})V_{tb}^{\ast
}V_{ts}~\epsilon_{\alpha\beta}^{tL}\ln\frac{\Lambda}{m_{w}}|Br(B^{+}%
\longrightarrow\rho^{0}l^{+}\nu_{l})
\]

\[
\epsilon_{\tau\tau}^{tL}\leq1.5
\]

For $B\rightarrow X_{s}\nu\overline{\nu}$ the Br and constraints are
\[
Br(B\rightarrow X_{c}\nu\overline{\nu})_{Int\operatorname{erf}erence}%
=\frac{\alpha_{em}^{2}}{16\pi^{2}\sin^{4}\theta_{W}|V_{cb}|}|V_{ts}%
V_{tb}X(x_{t})V_{ts}V_{tb}\epsilon_{\alpha\beta}^{tL}\ln\frac{\Lambda}{m_{w}%
}|\frac{\overline{\eta}}{f(z)\kappa(z)}Br(B\longrightarrow X_{c}l\nu_{l})
\]

\[
\epsilon_{\tau\tau}^{tL}\leq0.8
\]

\[
Br(K_{L}^{0}\longrightarrow\pi^{0}\overline{\upsilon}\upsilon
)_{Int\operatorname{erf}erence}=r_{iso}\frac{2\alpha_{em}^{2}}{|V_{ub}%
|^{2}16\pi^{2}\sin^{4}\theta_{W}}|V_{ts}^{\ast}V_{td}X(x_{t})V_{tb}^{\ast
}V_{td}~\epsilon_{\alpha\beta}^{uL}\ln\frac{\Lambda}{m_{w}}|Br(K^{+}%
\longrightarrow\pi^{0}e^{+}\nu_{e})
\]

\section{Discussion and results}

We study three processes $B^{+}\longrightarrow\pi^{+}\overline{\upsilon
}\upsilon,$ $B^{+}\longrightarrow K^{\ast+}\overline{\upsilon}\upsilon
~$(exclusive) and $B\longrightarrow X_{s}\overline{\upsilon}\upsilon$
(inclusive). which are theoretically clean processes. So, these are ideal for
the search of new physics. Very high $Br$ of $B^{+}\longrightarrow\pi
^{+}\overline{\upsilon}\upsilon$ and $B^{+}\longrightarrow K^{\ast+}%
\overline{\upsilon}\upsilon$ making it very attractive for the
experimentalists too. Although, $B\longrightarrow X_{s}\overline{\upsilon
}\upsilon$ is very difficult to detect but it is much clean as compared to any
other rare decay that's why it is studied for the search of new physics. The
results are summarized in table 1 and plots are provided in figures 3, 4,5,6,7
and 8 to make the comparison more clear. The constraints on NSIs with the
decays of charm and kaon rare decays involving neutrinos in their final states
are calculated in \cite{shakeel}\cite{shakeel 1}\cite{shakeel 2}, which give
very precise constraints $O(10^{-2})~$for the up type quarks. For the charm
and kaons the interference between the standard model and NSIs is very small,
$O(10^{-3})$ less than $NSIs$. But, for the case of $B$ rare decays having two
neutrinos in their final state the dominant contribution is coming from the
interference. One more difference is that for charm and kaons the Br of first
and second generation are giving higher values, but, for B the contribution
from second and third generation is providing leading contribution as compared
to the first generation. As for as the comparison with in the generation is
concerned we are getting more precise constraints with second generation of
quarks (c quark) for B, D and K rare decays having neutrino in final state.
But, the constrains from $B$ decays are less precise for $u$-quark than D and
K decays. It is contrary to the usual NSIs, in which the dominate contribution
comes from $u$-quark induced processes. The rare B decays will be in the range
of much clean environment of the B factories. That's why the analysis of such
reaction is very important for the new physics. We analysis $K_{L}%
^{0}\longrightarrow\pi^{0}\overline{\upsilon}\upsilon$ for the study of NSIs
with three generations of quarks which provides even be more precise
constraints as compared to D and K decays of same type. The constraint for
$K_{L}^{0}\longrightarrow\pi^{0}\overline{\upsilon}\upsilon$ are $O(10^{-3})$
for all the generations but the Br for third generations is very small and can
be ignored. This is shown in figures 9,10 and 11.%

\begin{figure}
\centering
%

{\includegraphics[
natheight=2.552100in,
natwidth=3.749800in,
height=2.5936in,
width=3.7974in
]%
{C: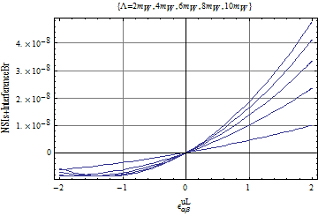}%
}
%

\caption{u quark induced NSIs of B+  Decays to Pi+, neutrino and antineutrino}
\end{figure}%
%

\begin{figure}
\centering
%

{\includegraphics[
natheight=2.531300in,
natwidth=3.749800in,
height=2.5728in,
width=3.7974in
]%
{C: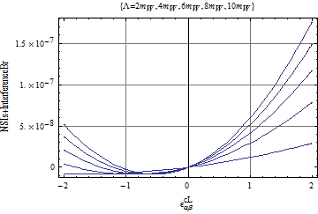}%
}
%

\caption{c quark induced NSIs of B+  Decays to Pi+, neutrino and antineutrino}
\end{figure}%
%

\begin{figure}
\centering
%

{\includegraphics[
natheight=2.531300in,
natwidth=3.749800in,
height=2.5728in,
width=3.7974in
]%
{C: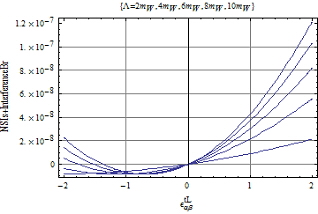}%
}
%

\caption{tquark induced NSIs of B+  Decays to Pi+, neutrino and antineutrino}
\end{figure}%
%

\begin{figure}
\centering
%

{\includegraphics[
natheight=2.479400in,
natwidth=3.749800in,
height=2.5201in,
width=3.7974in
]%
{C: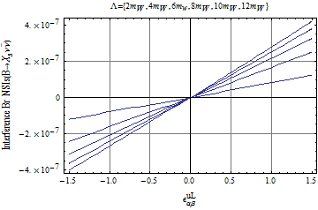}%
}
%

\caption{u quark induced NSIs of B Decays to Xc, neutrino and antineutrino}
\end{figure}%
%

\begin{figure}
\centering
%

{\includegraphics[
natheight=2.572800in,
natwidth=3.749800in,
height=2.6143in,
width=3.7974in
]%
{C: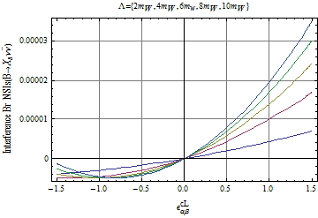}%
}
%

\caption{c quark induced NSIs of B Decays to Xc, neutrino and antineutrino}
\end{figure}%
%

\begin{figure}
\centering
%

{\includegraphics[
natheight=2.479400in,
natwidth=3.749800in,
height=2.5201in,
width=3.7974in
]%
{C: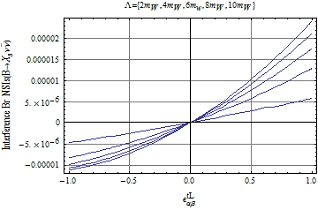}%
}
%

\caption{t quark induced NSIs of B Decays to Xc, neutrino and antineutrino}
\end{figure}%
%

\begin{figure}
\centering
%

{\includegraphics[
natheight=2.395500in,
natwidth=3.749800in,
height=2.437in,
width=3.7974in
]%
{C: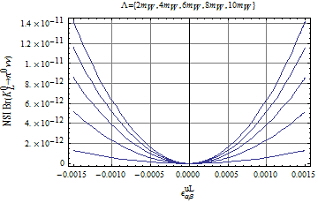}%
}
%

\caption{u quark induced NSIs of KL Zero Decay}
\end{figure}%
%

\begin{figure}
\centering
%

{\includegraphics[
natheight=2.395500in,
natwidth=3.749800in,
height=2.437in,
width=3.7974in
]%
{C: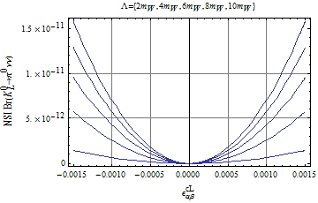}%
}
%

\caption{c quark induced NSIs of KL zero Decay}
\end{figure}%
%

\begin{figure}
\centering
%

{\includegraphics[
natheight=2.395500in,
natwidth=3.749800in,
height=2.437in,
width=3.7974in
]%
{C: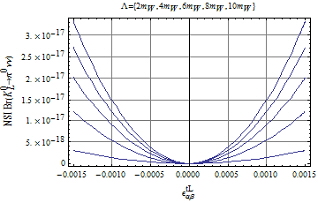}%
}
%

\caption{t quark induced NSIs of KL Zero Decay}
\end{figure}%
%

\begin{table}
\centering
%

\begin{tabular}
[c]{|l|l|l|l|l|l|}\hline
Reaction & Theoretical & Experimental & NSIs with $u$ & NSIs with $c$ & NSIs
with $t$\\\hline%
\begin{tabular}
[c]{l}%
$B^{+}\longrightarrow\pi^{+}\overline{\upsilon}\upsilon$\\
$(u\overline{b})\longrightarrow(\overline{d}u)\overline{\upsilon}\upsilon$%
\end{tabular}
& $\underset{\text{\cite{arxiv 2014}}}{1.5\times10^{-7}}$ & $\underset
{\text{\cite{PDG 2014}}}{<9.8\times10^{-5}}$ &
\begin{tabular}
[c]{l}%
$3\times10^{-8}$\\
$\epsilon_{\tau\tau}^{uL}\leq1.5$%
\end{tabular}
&
\begin{tabular}
[c]{l}%
$1\times10^{-7}$\\
$\epsilon_{\tau\tau}^{cL}\leq1.5$%
\end{tabular}
&
\begin{tabular}
[c]{l}%
$1\times10^{-7}$\\
$\epsilon_{\tau\tau}^{tL}\leq1.5$%
\end{tabular}
\\\hline
$%
\begin{tabular}
[c]{l}%
$B\longrightarrow X_{s}\overline{\upsilon}\upsilon$\\
$b\longrightarrow s\overline{\upsilon}\upsilon$%
\end{tabular}
\ $ & $3.6\times10^{-5}$ & $\underset{\text{\cite{ALEPH collaboration}}%
}{<64\times10^{-5}}$ &
\begin{tabular}
[c]{l}%
$\symbol{126}10^{-8}$\\
$\epsilon_{\tau\tau}^{uL}\leq0.8$%
\end{tabular}
&
\begin{tabular}
[c]{l}%
$1\times10^{-5}$\\
$\epsilon_{\tau\tau}^{cL}\leq0.8$%
\end{tabular}
&
\begin{tabular}
[c]{l}%
$1\times10^{-5}$\\
$\epsilon_{\tau\tau}^{tL}\leq0.8$%
\end{tabular}
\\\hline
$%
\begin{tabular}
[c]{l}%
$B^{+}\longrightarrow K^{\ast+}\overline{\upsilon}\upsilon$\\
$u\overline{b}\longrightarrow(u\overline{s})~\overline{\upsilon}\upsilon$%
\end{tabular}
\ $ & $3.56\times10^{-6}$ & $<4\times10^{-5}$ &
\begin{tabular}
[c]{l}%
$\symbol{126}10^{-8}$\\
$\epsilon_{\tau\tau}^{uL}\leq1.5$%
\end{tabular}
&
\begin{tabular}
[c]{l}%
$1.5\times10^{-6}$\\
$\epsilon_{\tau\tau}^{cL}\leq1.5$%
\end{tabular}
&
\begin{tabular}
[c]{l}%
$1.5\times10^{-6}$\\
$\epsilon_{\tau\tau}^{tL}\leq1.5$%
\end{tabular}
\\\hline
\end{tabular}
%

\caption{Comparison of the contstraints for B decays}
\end{table}%
%

\begin{table}
\centering
%

\begin{tabular}
[c]{|c|c|c|c|c|c|}\hline
Reaction & Theoretical & Experimental & NSIs with $u$ & NSIs with $c$ & NSIs
with $t$\\\hline
$\underset{(u\overline{b})\longrightarrow(\overline{d}u)\overline{\upsilon
}\upsilon}{K_{L}^{0}\longrightarrow\pi^{0}\overline{\upsilon}\upsilon}$ &
$2.06\times10^{-11}$ & $\underset{\text{\cite{PDG 2014}}}{<2.6\times10^{-8}}$
& $%
\begin{tabular}
[c]{l}%
$1.4\times10^{-11}$\\
$\epsilon_{\tau\tau}^{uL}\leq O(10^{-3})$%
\end{tabular}
$ & $%
\begin{tabular}
[c]{l}%
$1.5\times10^{-11}$\\
$\epsilon_{\tau\tau}^{cL}\leq O(10^{-3})$%
\end{tabular}
$ & $%
\begin{tabular}
[c]{l}%
$3\times10^{-17}$\\
$\epsilon_{\tau\tau}^{tL}\leq O(10^{-3})$%
\end{tabular}
$\\\hline
\end{tabular}
%

\caption{Comparison of the contstraints for KL}
\end{table}%

\section{Conclusion}

We have studied $B^{+}\longrightarrow\pi^{+}\overline{\upsilon}\upsilon
,B^{+}\longrightarrow K^{\ast+}\overline{\upsilon}\upsilon$, $B\longrightarrow
X_{s}\overline{\upsilon}\upsilon$ and $K_{L}^{0}\longrightarrow\pi
^{0}\overline{\upsilon}\upsilon$ for the search of new physics in the from of
"Non Standard Neutrino Interactions" (NSIs). We have calculated the Branching
ratio (Br) of these reactions and constrained NSIs by using the mismatch
between standard Model and the experiments. NSIs are giving very small
contributions in rare decays of $B$ mesons.as compared to the $D$ and $K$
decays involving neutrinos in their final state. But, the interference effects
for $B$ are giving dominant contribution and it is used to find out constraint
on NSIs. We found the contrarians for three generations of up type quarks.as;
$\epsilon_{\tau\tau}^{uL},\epsilon_{\tau\tau}^{cL}$ and $\epsilon_{\tau\tau
}^{tL}$ using interference.$~$Charm quark and top quark induced constraint
,$\epsilon_{\tau\tau}^{cL},\epsilon_{\tau\tau}^{tL}$ are much more precise as
compared $u$ quark. For $B^{+}\longrightarrow K^{\ast+}\overline{\upsilon
}\upsilon$ and $B\longrightarrow X_{s}\overline{\upsilon}\upsilon,$ the
$\epsilon_{\tau\tau}^{uL}$constraint is same but the Br is very small. This
shows that NSIs will have affects on the rare decays of $B$ meson both
inclusive and exclusive. $K_{L}^{0}\longrightarrow\pi^{0}\overline{\upsilon
}\upsilon$ is showing exactly the same behavior as of the $D$ and $K$ decays
but providing more precise bounds. Here the interference effects are small and
the bounds from $K_{L}^{0}\longrightarrow\pi^{0}\overline{\upsilon}\upsilon$
are $O(10^{-3})$ which are $O(10^{-1})$ smaller than $D$ and $K$.

\end{document}